# Prompt electrons driving ion acceleration and formation of a two temperatures plasma in nanosecond laser-ablation domain

D. Mascali[1], S. Tudisco[1,*], N. Gambino[1,2], A. Pluchino[2,3], A. Anzalone[1], F. Musumeci[1,2], A. Rapisarda[2,3], A. Spitaleri[1,2]

[1] INFN-Laboratori Nazionali del Sud, Via S. Sofia 62, I95123 Catania Italy
[2] Dipartimento di Fisica e Astronomia, Università di Catania, Via S. Sofia 64, I95123 Catania Italy
[3] INFN-Sezione di Catania, Via S. Sofia 64, I95123 Catania Italy

**Abstract** – We present the results of an experiment on plasma generation via laser ablation at $10^{12}$ W/cm$^2$ of power intensity and in a nanosecond domain. Prompt electrons emission and complex plasma plume fragmentation were simultaneously observed for the first time in this laser intensity regime, along with a double electron temperature inside the plasma bulk surviving for a long time to the plume expansion. 1D PIC simulations are in agreement with experimental data as long as the emission of initial prompt electrons is considered. This assumption results to be the key to explain all the other experimental evidences.

**Introduction.** – It is well known that high intensity lasers ($I>10^{13}$ W/cm2) interacting with solid targets generate plasmas with a hot-electrons (suprathermals) component [1-4]. Quite controversial and still under investigation [5-6] is the existence of this component for moderate intensity and long pulses regimes ($I<10^{13}$ W/cm$^2$, $\tau>0.1$ ns), due also to the diagnostics limits. At $I>10^{13}$ W/cm$^2$, the hot-electrons component was often invoked as the cause of complex ion acceleration mechanisms, leading to plume splitting or fragmentation [5,7].

Some authors [7] deduced that in the $I<10^{13}$ W/cm$^2$ - $\tau>0.1$ ns domain, and assuming a two-electron-temperature (TET) plasma, a rarefaction shock wave might be formed. It conceptually represents a double layer (DL), i.e. a region of non-neutral plasma that induces a large potential drop thereby causing the formation of very strong electric fields.

According to this model, shock-wave driven DL would be directly correlated with the two-temperature plasma, and should depend on the hot to cold electron temperatures ratio. In some experiments [5,10], the prompt-electrons component escaping from the interaction region has been observed and interpreted [5] as a signature of a TET plasma formation.

In this letter we report the results of an experimental campaign in which, for the first time, *i)* prompt electrons, *ii)* plume fragmentation in a multi-layers structure and *iii)* the existence of a double electron temperature inside the plasma bulk surviving to the plume expansion phase, have been simultaneously observed. In order to understand the fundamental underlying mechanisms in such observations, we performed 1D particle in cell (PIC) simulations which are in good agreement with experimental data.

**Experimental set-up.** – The experiment was carried out at Laboratori Nazionali del Sud, by using a Quanta System Nd:YAG laser, at 1064 nm of wavelength, 600 mJ of energy and 6 ns of pulse duration (FWHM). The laser beam was focused inside a vacuum chamber (where the operating pressure was kept constant at $4 \cdot 10^{-6}$ mbar) on a pure Aluminum target (2 mm thick) in order to get a power density of the order of $10^{12}$ W/cm$^2$; more details on the experimental set-up are reported in [11].

The plasma parameters were measured with a movable Langmuir Probe (LP) placed at different distances from the laser-target interaction area. The probe was located parallel to the expansion direction and biased from -100 up to +100 V to determine the plasma resistivity curve and to work as a time-of-flight (TOF) detector.

**Measurements.** – The first part of the experiment was carried out by using the unbiased LP as TOF measurement tool. We took care to collect signals from a totally polished target surface (a low intensity laser pulse was preliminarily launched on the target in order to remove the layers of possible contaminants). In the TOF mode, the probe particle collection was only governed by the self-consistent potential established inside the plasma sheath. The signals collected when the probe was located from 2.5 mm up to 14.5 mm from the target surface are reported in figure 1a. The negative part, highlighted in figure 1b and labelled as "*ph-e*", do not depend on the probe position: it exhibits the same temporal width of the laser pulse, that was fixed as zero time of the TOF spectra, and it is due to the photo-excitation of the probe tip. Figure 1b reveals that an additional negative excess of current at *d*=2.5 mm has been detected. This signal is due to a very fast prompt electrons bunch. In figure 1a we note also a "fragmentation" of the plasma signal: measurements feature primarily a series of positive bumps with decreasing amplitudes, characterized by an uncoupled dynamics with respect to the remaining part of the plasma, and then what we called the "plasma core". Plasma core expands at much lower velocity than bumps, as confirmed by the second part of the experiment, and its behaviour is well reproduced by hydro-dynamical simulations [11].

(*)E-mail: tudisco@lns.infn.it
(*)Present address: S. Tudisco Laboratori Nazionali del Sud – INFN Via S. Sofia 62, I95123 Catania Italy.

Further information was extracted by current-voltage characteristics (IVC) of the Langmuir probe. Collected signals at 0 and ± 60 V are reported in figure 1c, evidencing hot electrons contribution in the tail of the plasma core signal.

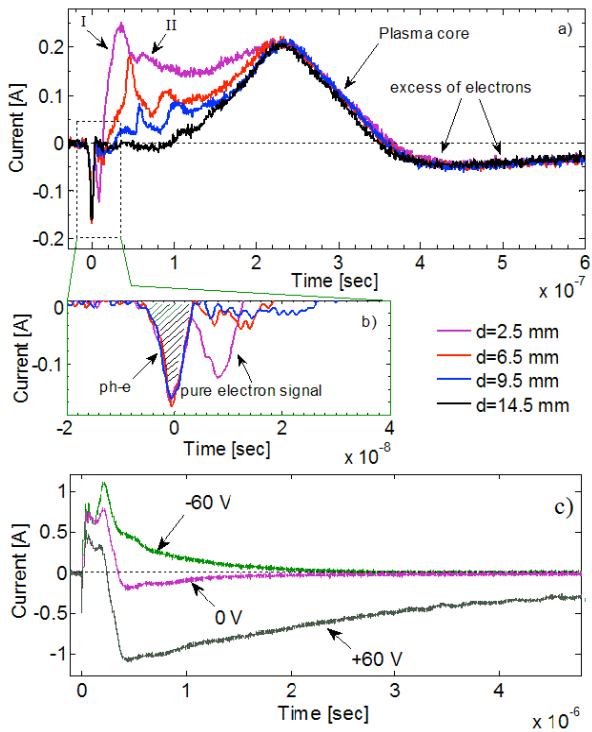

Fig. 1: (*a*) Time of Flight signals obtained when the probe was located from 2.5 mm up to 14.5 mm far from the target surface. (*b*) Prompt electrons signal. (*c*) Langmuir Probe Signals obtained at 4.5 mm from the target surface and at biasing voltages of ± 60 V.

**Data analysis.** – The IVC were analyzed as a function of time in order to extract time resolved electron temperature and density evolution. In the temporal windows up to 150 ns (i.e. the region where ions bunches were detected), the positive current did not vanish for any positive repelling probe potential; this evidence is a clear signature of an initial excess of ions, i.e. the plasma quasi-neutrality is violated inside the fast ions bunches; estimated average energies of ions which populate the bunches I and II are on the order of 1.5 ÷ 2 keV. Prompt electrons and ions bunches velocities as a function of time are reported in figure 2: the ions bunches emission is delayed with respect to prompt electrons and the ions acceleration proceeds on a timescale of hundreds of ns.

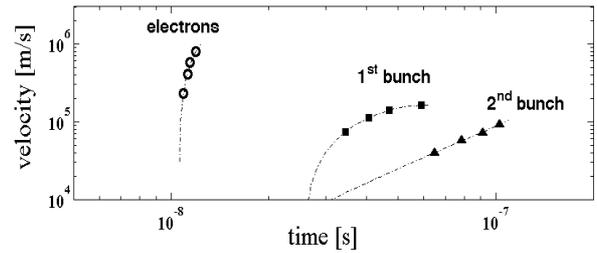

Fig. 2: Velocity vs. time plot of both prompt electrons and ions bunches. Dashed lines represent the results of 2nd order polynomial fits.

In the plasma core region the current-voltage curve exhibited the classical *S*–like behavior [12], as expected for quasi-neutral plasma. There the electron density and the electron temperature were calculated [12] according to classical probe theories [11, 13,14]. The estimated maximum density was $1.5 \cdot 10^{18}$ m$^{-3}$ at 14.5 mm far from the target. At 4.5 mm a two temperature plasma, with $T_e^{cold}$=3 eV and $T_e^{hot}$=17eV, was detected [11]. In a timescale of about 2 μs the two populations were thermalized, thus becoming undistinguishable. The tail of hot electrons coming at later times (up to several μs), that we observe in fig1c, was explained as due to the three-body recombination: by this mechanism, cold electrons can transfer energy to the hot ones, which in turn get an additional portion of energy and thus typically "survive" for longer times if compared to the cold ones [14].

The two-electron-temperature plasma (TET) has never been experimentally detected before in correlation with plume fragmentation. Actually, the existence of an "ab-initio" TET configuration was invoked by several authors [7,8,15] as the main source of plasma instabilities, especially in high intensity regime; several hypothesis have been suggested to explain the generation mechanism of the hot-electrons component (e.g. two-plasmon decay [16-18], stimulated Raman scattering [19], etc.). According to these models, the prompt-electrons should be made of particles populating the extreme tail of the initial Electron Energy Distribution Function (EEDF), i.e. they should be a fraction of the hot component escaping from the interaction region [5]; since they are collision-less, they do not undergo to any cooling during the expansion, thus keeping their own initial temperature.

However, this hypothesis was inconsistent with the estimation of the prompt-electrons temperature through a Gaussian fit of the TOF signals reported in fig.1b. In fact, we evaluated a temperature $T_e^{prompt} < 1$ eV, which is extremely low, even if compared with the hot plasma component detected at 4.5 mm far from the target (i.e. after cooling). This means that the prompt electrons are decoupled from the initial plasma bulk hot electron component. On the other hand, their mean velocity ($\sim 10^5$ m/s) is consistent with the quiver velocity $v_q=eE/m\omega_L$ in the laser field.

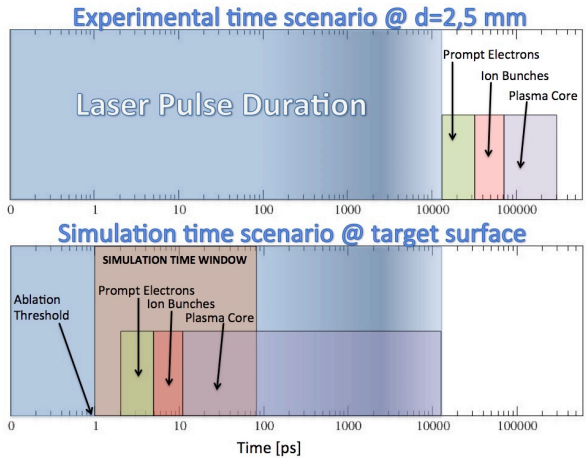

Fig. 3: Top panel: a pictorial view of the experimental time window with the main components (prompt electrons, ion bunches, plasma core) observed after the laser pulse at a distance of 2.5 mm from the target (see fig.1). Bottom panel: the same components are reproduced by the PIC simulation during the laser pulse (see fig.5) near the target surface (the simulation time window [80ps] is explicitly reported). In both the panels the intensity of the laser pulse is proportional to the intensity of the colour (the maximum intensity is reached after about 6 ns).

**Simulations.** – On the basis of these experimental evidences, we performed numerical simulations in order to better explore the roles of the prompts and eventually of the "ab-initio" TET hypothesis on the plasma evolution.

We used the particle-in-cell (PIC) method [21] over a 1D lattice. This method, based on mean field approximation, neglects individual close collisions between particles, while it is able to estimate the influence of the Coulomb interaction in the expanding plasma cloud. The plasma flow was treated as having basically 1D behaviour (more details are reported in [20]) and an idealised model of thermal evaporation from the target surface at constant temperature was inferred. In particular, a constant evaporation flux approximation was adopted and the evaporated particles were assumed to be instantaneously ionised. The evaporation follows a preformed plasma sheet extending on the lattice for several Debye lengths ($x_L \sim 10\lambda_D$ with $\lambda_D = 10^{-8}$m).

In figure 3 we present a pictorial view which compares, from a temporal point of view, the experimental scenario shown in fig.1 and obtained placing the probe at 2,5 mm far from the target surface (top panel), with the simulation scenario described in this section (bottom panel). Notice that the simulation time window covers the very early stage of the plasma dynamics, i.e. the first 80 ps immediately after the ablation threshold. In this way we expect to capture the original formation, within the plasma sheet, of the main components (prompt electrons, ion bunches, plasma core) experimentally observed at a greater distance. The main laser energy deposition, occurring much later (6 ns), can be studied only through a hydro-dynamical approach, as already done in ref. [11]. In any case, since it essentially affects the plasma core evolution, it is not relevant for our purposes.

In each simulation, the initial particles' space distribution for ions and electrons are given by two Fermi distributions:

$$F_e(x) = N_e[e^{(x-x_L)/d_e}+1]^{-1} \quad , \quad F_i(x) = N_i[e^{(x-x_L)/d_i}+1]^{-1}$$

where $N_e$ and $N_i$ are the initial number of electrons and ions in the plasma sheet ($N_e=z_iN_i$, where $z_i$ is the ion i-th charge state). The parameters $d_e$ and $d_i$ regulate the shape of the distribution around $x_L$, in a manner that electrons and ions are initially distributed in the intervals $[0, x_e^{max}]$ and $[0, x_i^{max}]$, with a spatial displacement $\delta=x_e^{max}-x_i^{max}$. Ions at different charge states have been included in the simulation, following a Gaussian-like distribution peaked on the value $<z>=4^+$. During the simulation, the initial quasi-neutrality is maintained only globally and not locally, by introducing the same amount of negative and positive charge into the 1D lattice. Finally, initial shifted-Gaussians (Maxwellian) velocity distributions for either electrons and ions in plasma sheet and plasma flow were assumed. Both the initial electron temperature ($T_e \sim 10^2$ eV) and the initial electron density ($n_e \sim 10^{25}$ m$^{-3}$) were extrapolated through hydro-dynamical simulations calibrated on the experimental data presented above. The mean velocity of the adiabatic expansion was estimated through the relation: $v=(\gamma KT_e/M_i)^{1/2}$ [11], where $M_i$ and $\gamma$ are the ion mass and the adiabatic expansion coefficient respectively. Finally, the total number of particles of the simulation, $N=N_e+N_i \approx 4\cdot10^5$, has been fixed by using the initial electron density while the time-step and the lattice-pitch were fixed in order to guarantee stable results and low numerical fluctuations ($\Delta t=0.02 \ \omega_{pe}^{-1}$, $\Delta x = 0.78 \ \lambda_D$).

In figure 4 we show the time evolution of a TET plasma: in the left column we report the particles' space distributions, with bins optimized for the visualization, while in the right column we plot the corresponding electrons velocity distributions. Plasma was prepared with $d_e=d_i=0.01 \ \lambda_D$, $x_L=10\Delta x$ and $T_e^{hot}=5T_e$ (see ref.[7]) at $t=0$ s (fig.4a and 4b). In this case the initial spatial distributions of both ions and electrons are step-like, therefore $x_e^{max}=x_i^{max}=x_L$ and $\delta=0$ (no displacement). We also assume an equal number of hot and cold electrons according to the experimental data [14]. In figure 4c, the expected hot electrons tail (but not a bunch!) escaping from the plasma reaches the extreme lattice boundary (L=$10^{-5}$ m) at t=2.14 ps, while the system is cooling down (fig.4d). Finally, at t=64 ps, no plume splitting (i.e. no multi-fragmentation) is observed even if the TET plasma still survives (fig.4e and 4f). Therefore the "ab-initio" TET plasma hypothesis (without displacement) seems not able to reproduce all the features of the experimental results (i.e. either generation of a prompt electrons bunch and plume fragmentation) although the large fraction of hot electrons fixed in initial conditions.

On the basis of the data reported in figure 2, which highlight a consistent time delay between prompts expulsion and ions bunches propagation, we realized new simulations assuming a non-neutral layer of electrons in the plasma front, which is virtually equivalent to time delay in the electrons emission. Only one electron temperature $T=T_e$ was set in this case.

evolution after 1.86 ps, featuring a bunch of fast escaping electrons whose dynamics are decoupled from the rest of the plasma cloud and with a velocity consistent with the experimental one; in figure 5d, correspondingly, a suprathermal electron component appears, whose distribution function deviates from the initial Gaussian shape. Finally, in figure 5e (t=80ps), it is evident that the expulsion of the prompt electrons has produced a non neutral positive layer in the plasma front, while the electron velocity distribution is now formed by two different components: one hot and the other one cold, corresponding to a double temperature plasma ($T_e^{hot}/T_e^{cold} \approx 2.5$ at 80 ps).

Figure 6 helps to better characterize the non-neutral positive layer moving at the plasma front. The 3D ions phase space evidences the strong acceleration of the higher charge states: they are characterized by larger velocity and placed ahead with respect with the lower charge states.

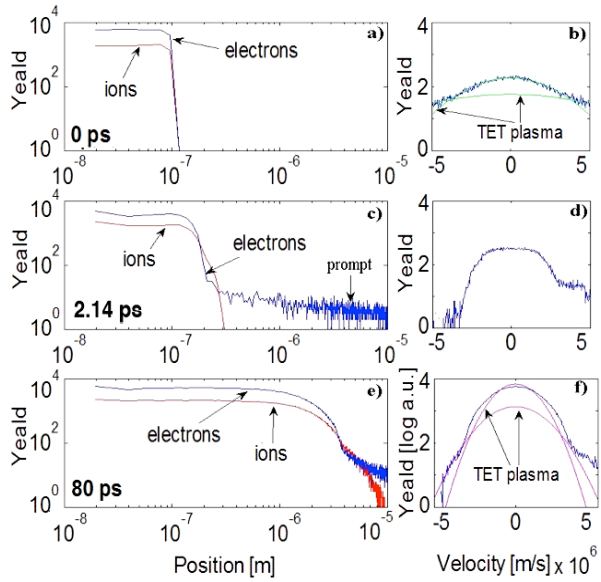

Fig. 4: PIC simulations on 1D lattice for an "ab-initio" TET plasma with no initial space displacement between ions and electrons. Snapshots of ions and electrons spatial distributions (on the left) and the corresponding electrons velocity distributions (on the right) at different time steps are displayed.

In figures 5 we show the time evolution of a plasma prepared with $x_L=10\Delta x$, $d_i=0.01\lambda_D$, $d_e=10\lambda_D$. We imposed a spatial cut-off at x=20Δx to the initial electrons distribution (Figure 5a), whose role is to fix the maximum extension of the preformed double-layer thus obtaining an initial displacement δ=10Δx, a value consistent with the literature [22].

The initial Gaussian velocity distribution of electrons with $T=T_e$ is displayed in figure 5b. Figure 5c shows the plasma

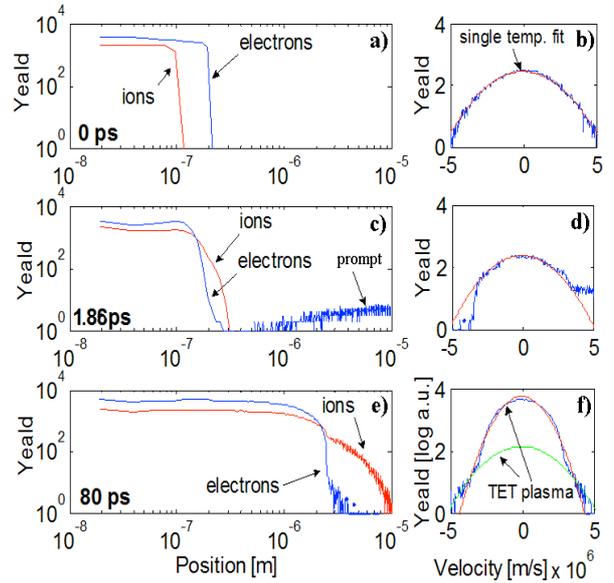

Fig. 5: PIC simulations on 1D lattice as in figure 3 but for a plasma with an initial spatial ions-electrons displacement and with only one initial temperature.

We then performed systematic simulations by setting different values of $d_e$, i.e. different shapes of the initial electrons space distribution, and always imposing the spatial cut-off at x=20Δx. In this way we tune the initial density of the electrons layer in the plasma front. For $d_e<4\lambda_D$ the cut-off starts to become unnecessary and there is no evidence of prompt electrons bunch formation, TET and plume fragmentation (electrons in the front are re-attracted by the positive slow-moving background). In the domain $4\lambda_D \leq d_e \leq 10\lambda_D$, first the prompt electrons bunch develops (for $4\lambda_D \leq d_e \leq 6\lambda_D$), and then the TET plasma and plume fragmentation evolve (in the range $6\lambda_D \leq d_e \leq 10\lambda_D$). The fragmentation is only initially driven by prompt electrons: their role is to leave a background of positive ion layers, which in turn explode because of self-repulsion. Then the ion acceleration proceeds because of reciprocal repulsion among the layers of different charge states, while the prompts flow freely moves being decoupled from the ions.

The above picture perfectly agrees with experimental data reported in Figure 2, which in fact show a slight delay between the expulsion of the prompt electrons and the emission of ionic bunches. Furthermore, it also confirms our initial expectation (see the bottom panel of fig.3) to observe numerically the original formation of the main experimental plasma components (prompt electrons, ion bunches and plasma core).

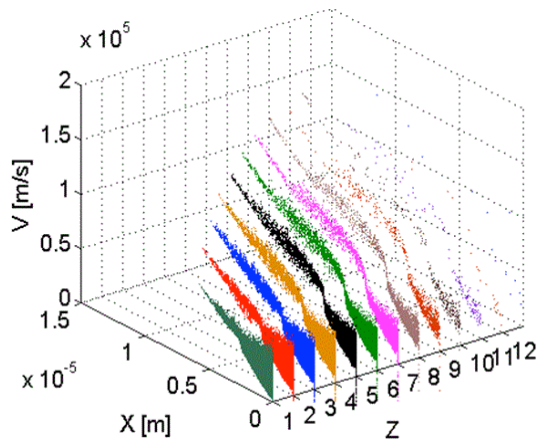

Fig. 6: PIC simulations. Ions phase space as a function of the charge state after 80 ps.

**Conclusions.** – On the basis of the showed results, the plasma evolution can be depicted in the following way: *i)* prompt electrons escape rapidly, but not before they have repelled part of the electrons in the plasma bulk; *ii)* if the initial density of electrons in the plasma front is large enough, i.e. prompt electrons are expelled well before the ion reaction time, of the order of $\omega_i^{-1}$, where:

$$\omega_i = qen_i^2/M\varepsilon_0 \ll \omega_e$$

the ions cannot compensate the compression of the bulk's electrons by attracting back the prompts (ions can react to the charge separation in a time of the order of tens of ps); while the prompt electrons flow at much larger velocity than the bulk plasma, so that they disappear from the simulation lattice, iii) a non neutral, positive layer in the plasma front develops. The reciprocal repulsion inside the positive layer leads to a non-linear fragmentation of the ion cloud in a plurality of positive bunches.

In conclusion, the synergy between experimental data and simulations shows for the first time that prompt electrons emission can directly trigger a complex plume fragmentation and a non-linear ions acceleration mechanism. The double electron temperature inside the plasma bulk, measured downstream, is the effect and not the cause of prompt electrons emission. Further investigations are necessary in order to understand the mechanism of prompt electrons formation and emission.

***

We acknowledge the 5th Nat. Committee of INFN for the financial support.